# Hidden Markov Model for Inferring Learner Task Using Mouse Movement

Elbahi Anis[#1], Mohamed Ali Mahjoub[*2], Mohamed Nazih Omri[#3]

*# Research Unit MARS,*
*Department of computer sciences*

*Faculty of sciences of Monastir, Monastir, Tunisia.*
[1]`Elbahi.anis@gmail.com`
[3]`MohamedNazih.Omri@fsm.rnu.tn`

*\* Research Unit SAGE,*
*National Engineering School of Sousse, Sousse, Tunisia.*
[2]`Medali.mahjoub@ipeim.rnu.tn`

*Abstract*— **One of the issues of e-learning web based application is to understand how the learner interacts with an e-learning application to perform a given task. This study proposes a methodology to analyze learner mouse movement in order to infer the task performed. To do this, a Hidden Markov Model is used for modeling the interaction of the learner with an e-learning application. The obtained results show the ability of our model to analyze the interaction in order to recognize the task performed by the learner.**

## I. INTRODUCTION

In the e-learning process, it is very important to know how the learner interacts with the interface. Indeed, analysis and recognition of the learner task may be too useful to provide feedback for the learning process in order to guide the learner.

The analysis of user behavior and activity recognition in a web user interaction process is one of the most popular subjects of the human computer interaction HCI and usability evaluation of web based applications.

Various studies have been achieved in this context to analyze the navigational behavior of the user in order to infer the activity provided using techniques such as eye tracking [1-3], psychological and physiological tracking [4] and mouse cursor tracking [5-8]. Many earlier works shows that it is obvious that Mouse trajectories can be processed, averaged, visualized, and explored for analyzing users' behaviors.

Generally, in e-learning the trajectory of the mouse is considered as a powerful tool to provide indicators of the way in which a learner interacts with an e-learning interface. In fact, mouse movements are guided by the goal of the task and reflect the cognitive processes of the user as shown in Fig. 1.

Have clues of cognitive processes of the learner using the trajectory of the mouse, can infer its goal, then know his needs and provide assistance for him in a feedback process.

In [9] "an inverse Yarbus process" whereby the authors infer the visual task by observing the measurements of a viewer's eye movements while executing the visual task.

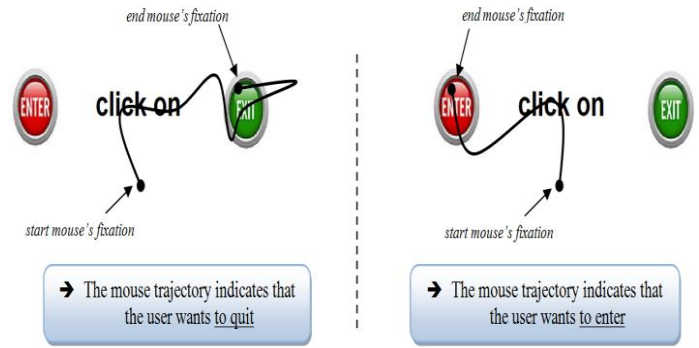

Fig. 1 Mouse trajectories for two different tasks

The use of eye tracking as a tool to infer the user task may not be an affordable alternative in various cases because the data provided by the eye tracking are expensive and may be biased, indeed subjects must be present in a specific location to wear or not tracking device and they are aware of the test situation hence the conditions of the experiment of gaze tracking may affect the found results much more than usual experience handling the mouse device.

Chen et al. [10] found that over 75% of cases, a mouse saccade moves to significant region of the screen and, in these cases, it is quite likely that the eye gaze is very near to the cursor.

In the present article, we propose to use the Markov theory to modelling the learner interaction using the cursor trajectory. The main goal is to infer the learner task in order to improve the e-learning process in a web based application.

## II. RELATED RESEARCH

The learner is the "center of gravity" of the e-learning process. In addition, adaptive and intelligent web-based educational systems attempt to be more adaptive by using the student profile associated with the goals, preferences and knowledge of the each individual student [15]. Various ways such as log files were used to better understand the progress of students and to guide them in the e-learning process [19,20], to provide good support for web users and to better understand the learner behavior. In the same way, the analysis of the trajectory of the mouse has been widely used to infer learner's strategy. Ohmori et al [14] analyzed the mouse movements during the reading task to make a classification of learners in three patterns. In [17] an application was suggested to track mouse movements of learners during their learning process in order to enable teachers to better understand the behavior of its students and in [18] the movements of the cursor were used to infer user interest over spatial information has shown to him. Various tools have been developed for the capture and analysis of the trajectory of the mouse like OGAMA[1] [13].

Freeman and Ambady [16] present "MouseTracker" tool that evaluate real-time processing in psychological tasks. Therefore the trajectory of the mouse has been widely used to guide and improve the process of learning since it reflects a part of the cognitive process of learner.

This type of interaction can potentially be used for the recognition of the task performed during the e-learning activity. Hence, this paper presents an HMM which is based on the path of the mouse to recognize the task performed by a learner in order to enhance the e-learning process.

III. HIDDEN MARKOV MODELS (HMM) :

*A. Presentation:*

Although HMM have been introduced since the late 60s by Baum and al, they are still overused in various fields for modelling stochastic sequences [12,9] as the analysis and speech recognition, handwriting recognition, image recognition, DNA sequence analysis, activity recognition, ...

In the present article, we do not try to detail the HMMs. However, for An excellent tutorial covering the basic HMM technologies the reader is recommended to see the documents of R. Rabiner [11,12]. In what follows we use the same notation used by R. Rabiner.

*B. Formal definition:*

Briefly, a HMM is a statistical mathematical model used to describe a doubly stochastic process, which can be presented by the following figure:

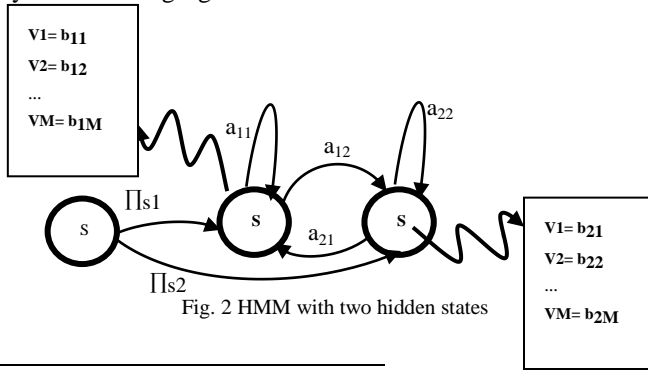

Fig. 2 HMM with two hidden states

[1] OGAMA available on : http://www.ogama.net/

Formally, a HMM noticed by $\lambda = (A, B, \Pi)$ is defined by:
- $S = \{S_1, S_2, ..., S_N\}$: set S of N hidden states with $S_0$ is a start state. (At time $_t$ the model state is $q_t$).
- $O = \{O_1, O_2, ... O_M\}$ : the alphabet of M symbols.
- $A = \{a_{ij}\}$ / $a_{ij} = p(q_t=S_j \mid q_{t-1}=S_i)$, $1 \leq i,j \leq N$, $\sum_{j=1}^{N} a_{ij} = 1$ : the matrix of transition probability between the N states.
- $B = \{b_j(k)\}$ / $b_j(k) = p(O_k = O_t \mid q_t = S_j)$, $1 \leq j \leq N$, $1 \leq k \leq M$, $\sum_{j=1}^{N} b_j(K) = 1$ : the matrix of symbol transmission probability by N states.
- $\Pi = \{\Pi_i\}$ / $\Pi_i = p(q_1=S_i)$, $1 \leq i \leq N$, $\sum_{i=1}^{N} \Pi_i = 1$ : the Initial probability distribution vector.

The system starts with an initial transition from state $S_0$ to all other states $S_i$ of the model with an initial transition probability $\Pi_i$ ; $0 \leq \Pi_i \leq 1$.

The system can pass from any state $S_i$ to an other state $S_j$ with a transition probability $a_{ij}$ ; $0 \leq a_{ij} \leq 1$.

From each state $S_j$ the system generates a symbol $O_k$ with an emission probability $b_j(k)$ ; $0 \leq b_j(k) \leq 1$.

*C. Markov property:*

A stochastic process has the Markov property if the prediction of the next state depends only on the relevant information contained in the present state of the process, ie:
$$P(X_{n+1}=j \mid X_0, X_1, ..., X_{n=i}) = P(X_{n+1}=j \mid X_{n=i})$$

*D. Problems and their solutions for HMM's :*

Like shown in TABLE I, Rabiner and Juang in [11] explain the three key problems of interest that must be solved for the model to be useful in real world application, given a model $\lambda = (A, B, \Pi)$ and observation sequence $O=\{O_1, O_2, ..., O_T\}$.

TABLE I
PROBLEMS / SOLUTIONS FOR HMM'S

| | Description of the problem | Solution algorithm |
|---|---|---|
| Prob1 | How we compute $P(O \mid \lambda)$. | Forward-backward |
| Prob2 | How we choose a state sequence $S=\{S_1, S_2, ..., S_T\}$ witch is optimal in some meaningful sense. | Viterbi |
| Prob3 | How we adjust the model parameters to maximize $P(O \mid \lambda)$ | Baum-Welch |

*E. HMM for task learner modelling.*

A task is the outcome of the interaction of the learner with the interface areas to achieve a fixed goal.

Since the handling of the interface areas is related to the aim of the task, the parts of the interface are not manipulated in the same manner for different tasks as shown in Fig. 1. Also, task can be illustrated as a set of states among them the system makes a transition at each time $_t$ with a certain probability. The transition to a state randomly generates an observable symbol which is the zone crosses by cursor. So, the task can be defined as a doubly stochastic process, the first stochastic level is the transition from one state to another randomly, and the second stochastic level is the random generation of a symbol which is the area of interest manipulated by the user cursor. Then a task user can be modelled using a HMM. In the next section we describe the proposed model.

## IV. PROPOSED MODEL

An e-learning activity can be defined as a set of task to perform. The identification of the task performed by the learner can significantly improve the e-learning process.

TASK={$TSK_1$, $TSK_2$, … ,$TSK_N$}, a set of task that can be performed by a user and each task is a set of actions.

To perform a task, the user can focus (fixes with mouse cursor) areas in the interface more than others; in fact the probability of using areas is strongly related to the task.

Let AOIs = {$Z_1, Z_2,…, Z_M$} the set of M areas of interest that can be manipulated or focused by the mouse during the task. Identifying AOIs can be done by an expert who fixed parts of the interface that are needed to perform all tasks required by the user. Each AOI is used to perform an action of the task.

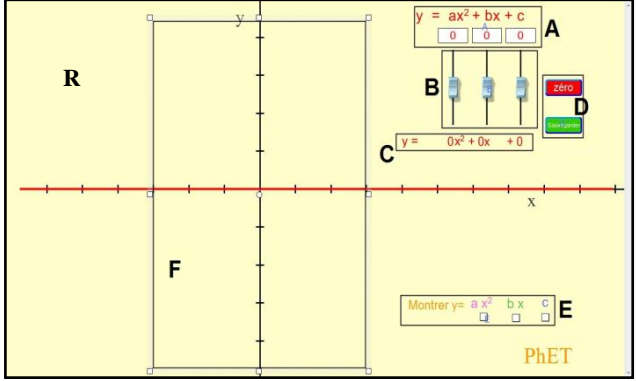

Fig 3: AOIs for the phet[2] "Equation Grapher" simulator.

A task is represented in our model by the trajectory of the mouse when a subject crosses a set of AOIs for a T period to achieve a goal. Hence, a task can be defined as follows:
TSK = {$Z_i(t)$}    $1 \leq i \leq M$ ; $1 \leq t \leq T$.

After recording the data of experience, a step of vectorization will take place. The purpose of this step is to identify all areas that have been crossed during the task using the following algorithm.

---

**Vectorization algorithm**:

*Begin vectorization*
*Initializations*
   O={ } ;
   Details of each area AOIs;
   T ← Total duration of the task performed;
   ds ← Time between recordings of tow cursor coordinates;
*Treatment*:
for t:=1 to T (with ds step) do
if (cursor coordinates of a mouse trajectory is in $Z_i$) then
   O [t] ← $Z_i$
endif
end for
*Output:*
   O = {$O_1, O_2, ..., O_T$}
*End vectorization*

---

[2] Phet available on http://phet.colorado.edu/sims/equation-grapher/equation-grapher_fr.html

The vectorization step result is O={$O_1, O_2, …, O_t, …, O_T$} ; $O_t = Z_i(t)$, $1 \leq i \leq M$, $1 \leq t \leq T$.

In order to infer the task achieved by the learner, we propose to use for each task, an HMM and we proceed as follows:

1 - Each task $TSK_k$ must be modeled by a specific HMM $\lambda_k = (\Pi, A, B)$ ; $1 \leq k \leq N$

2 - The parameters for each HMM $\lambda_k$ must be estimated, so we must train it by a set of observation sequences regarding the task $TSK_k$.

3 - For a task TSK caracterized by the sequence O = {$O_1, O_2, ..., O_t, ..., O_T$} and for each HMM $\lambda_k$, we must calculate the probability $P_{TSK} = P(O | \lambda_k)$ ; $1 \leq k \leq N$.

4 - To infer the task we choose the task whose probability value founded by the model is maximum, ie:

$$TASK^* = \underset{1 \leq k \leq N}{argmax} [P(O | \lambda_k)]$$

The following illustration shows the task inference process which is based on the calculation of maximum probability value generated by each HMM for the task performed.

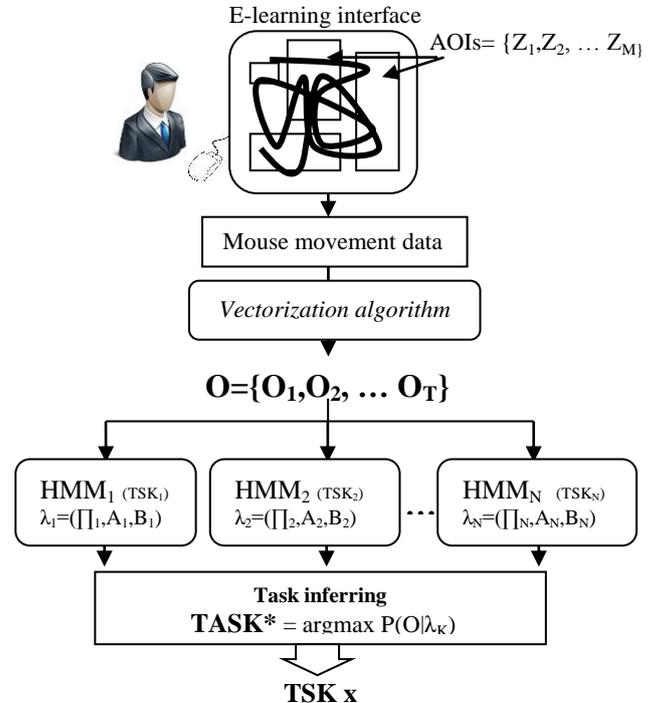

Fig 4. Task inference model using mouse movement data.

## V. EXPERIMENTS

In this section we will describe our proposed method.

On the Phet simulator, we fixed the details of interesting areas (AOIs) that can be manipulated by the learner using the mouse, as shown in Fig. 3. Then AOIs ={A,B,C,D,E,F,R} with R is the position of the mouse outside the areas A, B, C, D, E and F. For simplicity, we asked the students to achieve only one of the two tasks bellow.

$TSK_1$ = graphical representation of a curve (REP).

$TSK_2$ = graphic verification of the intersection of two curves (INT).

For modelling each task, we propose an initial model $\lambda = (\Pi, A, B)$ Which parameters are:

$$\Pi = \begin{pmatrix} 1 \\ 0 \end{pmatrix} \qquad A = \begin{pmatrix} 0.5 & 0.5 \\ 0.5 & 0.5 \end{pmatrix}$$

$$B = \begin{pmatrix} 0.3 & 0.2 & 0.05 & 0.1 & 0.05 & 0.2 & 0.1 \\ 0.2 & 0.1 & 0.05 & 0.15 & 0.3 & 0.1 & 0.1 \end{pmatrix}$$

For adjusting the parameters of each HMM, we prepared a training dataset for each hmm. Using the OGAMA tool [13] and the "equation grapher" simulation, we asked 10 participants to perform graphical equation representations of their choice. Similarly, we asked 10 others participants to perform a task of checking the intersection of two curves of their choice. Using Baum-Welch algorithm the, parameters obtained are:

$\lambda 1 = (\Pi 1, A1, B1)$ learned with the Learning Base (LB1) concerning the first task.

$$\Pi 1 = \begin{pmatrix} 1 \\ 0 \end{pmatrix} \qquad A1 = \begin{pmatrix} 0.9535 & 0.0465 \\ 0.0604 & 0.9396 \end{pmatrix}$$

B1=
$$\begin{pmatrix} 0.3818 & 0.3596 & 0.0379 & 0.0091 & 0.0046 & 0.0553 & 0.1517 \\ 0.0192 & 0.0096 & 0.0048 & 0.8827 & 0.0645 & 0.0096 & 0.0096 \end{pmatrix}$$

$\lambda 2 = (\Pi 2, A2, B2)$ learned with the Learning Base (LB2) concerning the second task.

$$\Pi 2 = \begin{pmatrix} 1 \\ 0 \end{pmatrix} \qquad A2 = \begin{pmatrix} 0.9509 & 0.0491 \\ 0.0501 & 0.9499 \end{pmatrix}$$

B2=
$$\begin{pmatrix} 0.2356 & 0.3419 & 0.0536 & 0.0092 & 0.0046 & 0.1764 & 0.1787 \\ 0.0183 & 0.0092 & 0.0046 & 0.0963 & 0.8533 & 0.0092 & 0.0092 \end{pmatrix}$$

Once the parameters of each HMM were estimated, 10 experiments were performed in which participants were asked to perform a task of their choice (REP / INT) and for each task an observation sequence $O = \{O_1, O_2, ..., O_T\}$ was generated using the vectorization algorithm cited above.

To infer the task performed, each obtained observation sequence is estimated using both HMM. The maximum likelihood value generated using Forward-backward algorithm will be considered as an indicator of the task performed by the participant. The results of the model are shown in next section.

## VI. RESULTS AND DISCUSSION

Using OGAMA [13], we record the path of the mouse and the heat map. The heat map can tell us about the most used and the most ignored areas in the interface, and the mouse path presents the order by which the learner interacts with interface simulator for each task performed as shown in the Fig. 4.

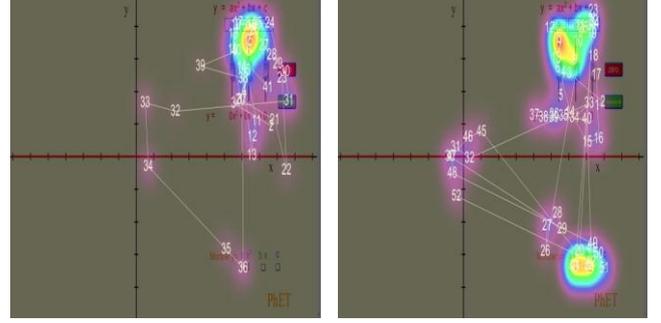

Fig 4. Two heat maps (with mouse paths) for two different tasks performed by the same learner using Phet simulator.

The examination of each heat map shows that -without being aware-, the learner draws a mouse path, fixes areas more than others and spends a lot more time on elements more than others. Despite the significance and quality of data provided by the heat map and the mouse path, it is difficult to automatically infer the type of task performed.

For the 10 experiments performed, the table II shows for each task, its duration, nature, percentage of mouse fixations on each area of interest, the logarithm of the likelihood probability generated by each HMM and decision taken by the model.

Remember that the HMM1 is part of the model trained to recognize TSK1 (REP) while HMM2 is part of the model trained to recognize TSK2 (INT).

For experiments T1, T2, T3, T5 and T9 in which participants made a simple graphical representation of the curve, the HMM1 generated higher values of probability then those generated by the HMM2.

While for experiments T4, T6, T7, T8 and T10 where participants checked the curves intersection, probability values generated by the HMM2 are higher than those generated by the HMM1.

TABLE II
EXPERIMENTAL RESULTS

| Task ID | Task time | Task type REP | Task type INT | Fixation % on each AOI A | B | C | D | E | F | R | HMM1(R) | HMM2(I) | DECISION |
|---|---|---|---|---|---|---|---|---|---|---|---|---|---|
| T1 | 2800 | | + | 0,57 | 0,39 | 0,50 | 2,11 | **45,32** | 14,82 | 36,29 | -6908.1 | **-3110.8** | INT |
| T2 | 3000 | | + | 0,00 | 8,00 | 0,70 | 1,70 | **70,60** | 5,43 | 13,57 | -7601.6 | **-1977.4** | INT |
| T3 | 3000 | | + | 0,00 | 0,33 | 0,17 | 0,00 | **88,77** | 1,67 | 9,07 | -8203.4 | **-1195.5** | INT |
| T4 | 1650 | + | | 36,42 | **43,70** | 3,52 | 7,58 | 0,00 | 0,06 | 8,73 | **-1884.5** | -2443.5 | REP |
| T5 | 3700 | | + | 12,89 | 28,49 | 3,22 | 7,54 | **30,16** | 6,24 | 11,46 | -6733.4 | **-4366.9** | INT |
| T6 | 2400 | + | | 15,67 | **51,13** | 7,42 | 7,88 | 3,25 | 5,54 | 9,13 | **-3379.6** | -3591.7 | REP |
| T7 | 2540 | + | | 26,14 | **36,18** | 6,34 | 8,23 | 3,15 | 4,92 | 15,04 | **-3579.6** | -3943.3 | REP |
| T8 | 2860 | + | | 27,27 | **34,20** | 4,79 | 15,63 | 4,72 | 3,25 | 10,14 | **-3618.4** | -4483.3 | REP |
| T9 | 2050 | | + | 16,10 | 17,80 | 3,41 | 5,37 | **47,46** | 0,00 | 9,85 | -4115.1 | **-1957.5** | INT |
| T10 | 2920 | + | | **44,73** | 34,04 | 0,24 | 9,69 | 0,00 | 3,56 | 7,74 | **-3223.8** | -4374.6 | REP |

So the model developed could infer the type of each task correctly. The following graph shows the capacity of each HMM of the model to recognize each task.

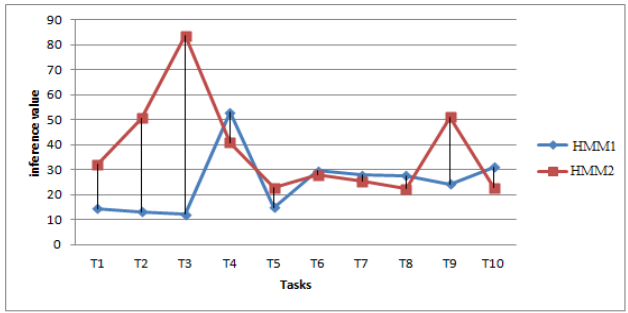

Fig 5. Task inference value for each HMM of the model

For experiments T1,T2,T3 and T9 there is a strong difference between the two HMMs whereas the other experiments the difference between the two HMMs is low.

Consider for example the experience T6 where the value generated by HMM1is -3379.6 and by HMM2 is -3591.7. Although the model could infer the task correctly, the difference between values founded by the two HMMs is very low (182,1), whereas for the experience T3 the difference between values founded by the two HMMs is higher (7007,9)

This can be explained by the fact that learner performing the task T6 focuses on the areas A (15,67%) and B (51,13%) which are recognized by the first HMM1 as very important areas for the realization of task REP (according to the emission matrix estimated for HMM1) but he manipulates other areas in relation to the task INT (C=7,42% ; D=7,88% ;E= 3,25% ;F=5,54%) , so the difference between tow HMMs is not very important.

Whereas in the case of task T3, the learner fixes more the area E (88,77%) which is considered very important for the realization of the task INT (according to the emission matrix estimated for HMM2) but he is not frequently uses areas in relation to the task REP (A=0% ;B=0,33% ;C= 0,17% ;D=0%; F=1,67%), so the difference between two HMMs is very important. The results of all the experiments can be explained in the same way.

From the other side, the inference value can give an indication on how learners interact with the e-learning application. In fact, for a learner doing a task, if the difference between two HMMs is important we conclude that the user targets well the areas needed to the achievement of the required task and therefore he has a good level of manipulation of the application.

But if learner performs the same task and the values returned by the two HMMs are too close we concluded that he has a low level of manipulation of the application than the first learner, so this learner presents problems of handling the application.

Another result that can be concluded from this work, according to the following graph which presents the rate of use of each area for the realization of all tasks .

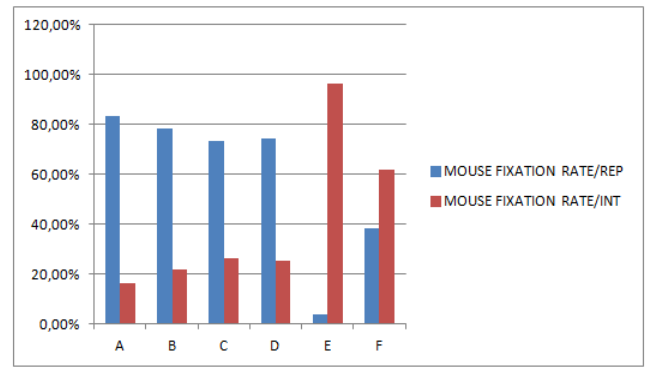

Fig 6. Mouse fixations rate in each AOI for each task.

The figure shows that the realization of the task INT attracts many the mouse cursor of the user on the area E and for the task REP, mouse cursor will be attracted to areas A and B. This results, brings us to conclude that the mouse movements are guided by the objective of the task to well-defined areas of the interface. This finding, is in accordance with to the work of Yarbus [19] who found that the gaze path of a subject is dependent of his task and with the work of Chen et al that show that the gaze path is strongly correlated with the mouse path[10].

## VII. CONCLUSIONS

In this paper we propose a model for learner interaction with an e-learning web based application.

The proposed model has good ability to infer the task performed by the learner based on mouse path and using a Hidden Markov Model for each task. The inference technique is based on the greater likelihood probability value generated by the HMMs of the model.

The model can give indications about the learner level of interaction with the application to help the teacher to know the problems faced by learners.

Despite their power and their wide use in various fields, HMMs have some problems like a choice of initial parameters of the model that may influence the effectiveness of the model, and also the training set which must be large and containing sequence learning well chosen. The enhancement of these elements can significantly improve the proposed model.